\documentclass[aps,prb,amsmath,amssymb,reprint]{revtex4-1}
\usepackage{epsfig,bm}

\begin{document}

\title{Elastic aging from coexistence and transformations of ferroelectric and antiferroelectric states in PZT}

\author{F. Cordero}
\email{francesco.cordero@isc.cnr.it}

\author{F. Craciun}
\affiliation{CNR-ISC, Istituto dei Sistemi Complessi,
Area della Ricerca di Roma - Tor Vergata,\\
Via del Fosso del Cavaliere 100, I-00133 Roma,Italy}

\author{F. Trequattrini}
\affiliation{Dipartimento di Fisica, Universit\`{a} di Roma
\textquotedblleft La Sapienza\textquotedblright , P.le A. Moro 2,
I-00185 Roma, Italy and CNR-ISC}

\author{P. Galizia}
\author{C. Galassi}

\affiliation{CNR-ISTEC, Istituto di Scienza e Tecnologia dei
Materiali Ceramici, Via Granarolo 64, I-48018 Faenza, Italy}

\date{\today}

\begin{abstract}
Materials undergoing antiferroelectric/ferroelectric (AFE/FE) transitions
are studied for possible applications that exploit the large volume, charge
and entropy differences between the two states, such as electrocaloric
cooling, energy storage, electromechanical actuators. Though certain
compositions of PbZr$_{1-x}$Ti$_{x}$O$_{3}$ (PZT) codoped with La and Sn may
withstand millions of electrically induced AFE/FE cycles, in other cases few
thermally induced cycles and room temperature aging may cause noticeable
changes in the material properties. This is particularly evident in the
elastic moduli, which at room temperature can become as much as four times
softer. In order to get more insight into the mechanisms involved in such
elastic aging and full recovering with mild annealing at $600-800$~K, the
effect of La doping on PbZr$_{0.954}$Ti$_{0.046}$O$_{3}$ is studied with
anelastic measurements. Complete suppression of the time dependent
phenomena is found after the transformation of the intermediate FE phase into
incommensurate AFE by 2\% La doping. This is discussed in terms of disappearance
of the stress and electric fields at the FE/AFE interfaces, in the light of the
thermally activated anelastic relaxation processes that are observed at high
temperature, and are due to mobile defects, presumably O vacancies.
\end{abstract}



% 77.80.B- Phase transitions and Curie point
% 77.84.Cg PZT ceramics and other titanates
% 62.40.+i Anelasticity
% 77.65.-j Piezoelectricity and electromechanical effects

\maketitle

\section{Introduction}

Among the antiferroelectric materials, PbZrO$_{3}$-based perovskites
have been intensively investigated due to their electrically induced
antiferroelectric/ferroelectric (AFE/FE) transition, which is
generally accompanied by large changes in volume and
charge\cite{HZK14}. These properties are very appealing for
potential use in various applications such as strain actuators, high
energy storage capacitors, pulsed power generators, \textit{etc}
\cite{HZK14,TMF11}. Various doping strategies have
been employed to tailor the phase transformation behavior of PbZrO$_{3}$%
-based materials in order to improve their performances. Thus, the most
widely used dopants to reduce the critical field are La$^{3+}$, Ba$^{2+}$, Ca%
$^{2+}$ on the Pb$^{2+}$ site and Sn$^{4+}$, Ti$^{4+}$ and Nb$^{5+}$
on the Zr$^{4+}$ site \cite{HZK14,HZZ15}. The AFE/FE transition in
PbZrO$_{3}$-based perovskites is also associated with high entropy
changes and electrocaloric/pyroelectric effects, which are very
interesting for novel applications such as solid
state cooling and pyroelectric energy harvesting \cite{MZS06,HYX11,FRK12,TS15}%
. For example, in Pb$_{0.97}$La$_{0.02}$(Zr$_{0.95}$Ti$_{0.05}$)O$_{3}$ a
maximum reversible adiabatic temperature change of 8.5~${{}^{\circ }}$C has
been obtained near the phase transition temperature together with a very
large recoverable energy density of 12.4~J/cm$^{3}$ at room temperature \cite%
{HYX11}. Materials that are both AFE and FE and show electrocaloric effect
and devices based on them have been recently reviewed in Refs. \onlinecite%
{FRK12,TS15,MKM14}, and their readiness for use in heat pumping
applications has been pointed out. Actually solid-state cooling
devices, including those based on the electrocaloric effect, are
among the priorities in various national and international energy
programs, as listed \textit{e.g.} in Refs. \onlinecite{FRK12,TS15}.
However, major challenges for introducing ferroic materials in
practical cooling applications are still unsolved. Since
electrocaloric cooling uses the entropy changes associated with
first order phase transformations, structural changes and fatigue
under repetitive electrical stress are important issues
\cite{FRK12}. Cycle stability and fatigue are also fundamental
issues for electromechanical actuators. Both refrigeration devices
and strain actuators must withstand a huge number of cycles during
their lifetime, and therefore understanding effects like creep and
fatigue, occurring at longer functioning time scale, is mandatory
for practical applications.

Previous works showed that the formation of the AFE phase in undoped
and doped PbZrO$_{3}$\ may be slow and accompanied by aging
phenomena, and it has been proposed that this is due to extended
defects forming at the FE/AFE interfaces in order to accommodate the
large volume difference between the AFE and FE phases \cite{PP99}.
Aging and thermal cycling through the AFE/FE transition may also
cause outstanding softenings of the Young's modulus and splitting of
the antipolar and octahedral tilt modes involved in the AFE/FE
transition, which however can be fully recovered with mild annealing
\cite{CCT13,CTC14}$^,$\footnote{In Refs. \onlinecite{CCT13,CTC14} we
identified the stiffening on cooling with the orthorhombic tilt mode
(OT) and the softening with the antiferroelectric mode (AF). At that
time, as only explanation for a steplike stiffening during cooling
through a structural phase transition we found the loss of one tilt
system from the disordered tilted phase (with octahedral rotations
about all axes) to the orthorhombic phase, untilted along the $c$
axis. Now we are convinced of the importance of the piezoelectric
softening in the FE phase, and therefore attribute the stiffening on
cooling to the AF mode and the softening to the OT mode. What is
written in Refs. \onlinecite{CCT13,CTC14} is valid if one exchanges
AF with OT everywhere, except for some inessential comments to few
fittings. Some of the fits of the $s^{\prime }\left( T\right) $
curves are slightly different, due to the compatibility conditions
between the different phases embedded in the fitting equations, but
the parameters remain identical or very similar with AF and OT
exchanged. Also the discussion of the effect of the loss of tilting
about the $c$ axis remains valid, though evidently it is not enough
to transform into stiffening the expected softening at the onset of
the long range order below $T_{\mathrm{OT}}$.}. In order to better
understand the nature of these phenomena in PbZrO$_{3}$-based
materials, and in particular whether the orthorhombic AFE phase
itself plays a particular role, we study here the effect of La
doping on the elastic aging. Doping with La destabilizes the FE
phase at the expenses of the AFE, and above 1\% La changes the
intermediate FE phase into incommensurate (IC) AFE
\cite{DXV95b,AK04}. Such a change completely suppresses the
irreversible softenings, that occur during aging and temperature
cycling, indicating that they are indeed due to the coexistence of
the AFE and FE phases.

\section{Experimental}

Ceramic samples of PbZr$_{0.954}$Ti$_{0.046}$O$_{3}$ (PZT 95.4/4.6
or PZT) and Pb$_{0.97}$La$_{0.02}$Zr$_{0.954}$Ti$_{0.046}$O$_{3}$
(PLZT 2/95.4/4.6 or PLZT) were prepared by the mixed-oxide method in
the same manner as previous series of samples \cite{127,145}. The
oxide powders were calcined at 800~${{}^{\circ }}$C for 4 hours,
pressed into bars for the anelastic and into pellets for the
dielectric experiments, sintered at 1250~${{}^{\circ }}$C for 2~h
and packed with PbZrO$_{3}$\ + 5wt\% excess ZrO$_{2}$\ to prevent
PbO loss. The powder X-ray diffraction did not reveal any trace of
impurity phases. The densities were about 95\% of the theoretical
values and the grains were large, with sizes of $5-20$~$\mu $m.
Diffractograms were also measured around the AFE/FE transition and
fitted with the Rietveld method, in order to extract the pseudocubic
cell volumes of the coexisting O-AFE ($Pbam$ space group
\cite{CGD97}), IC-AFE (no space group is reported for this phase and
we used $Pnnm$ \cite{LT99b}) and R-FE phases ($R3m$ space group
\cite{MMA69}).

For the anelastic experiments bars were cut 0.6~mm thick and up to
4~cm long and Ag paint electrodes were applied to the bars and
pellets. The dielectric permittivity $\varepsilon =$ $\varepsilon
^{\prime }-i\varepsilon ^{\prime \prime }$, with losses $\tan \delta
=\varepsilon ^{\prime \prime }/\varepsilon ^{\prime }$, was measured
with a HP 4284A LCR meter with a four-wire probe and an electric
field of 0.5 V/mm, between 0.2 and 200~kHz, controlling temperature
with a modified Linkam HFS600E-PB4 stage. The dynamic Young's
modulus $E=$ $E^{\prime }-iE^{\prime \prime }$, was measured by
suspending the bars on thin thermocouple wires in vacuum and
electrostatically exciting their flexural modes \cite{135}. The data
are presented as compliance $s=$ $1/E=$ $s^{\prime }-is^{\prime
\prime }
$, normalized to the stiffer value $s_{0}$ in the PE phase, with losses $%
Q^{-1}=s^{\prime \prime }/s^{\prime }$. The real part was deduced
from the resonance frequency $f\propto \sqrt{E^{\prime }}$
\cite{NB72}, as $s\left(
T\right) /s_{0}\simeq $ $f_{0}^{2}/f^{2}\left( T\right) $, with $f_{0}=1-5$%
~kHz for the fundamental mode, depending on the sample composition and
length.

\section{Results}

\begin{figure}[htb]
\includegraphics[width=8.5 cm]{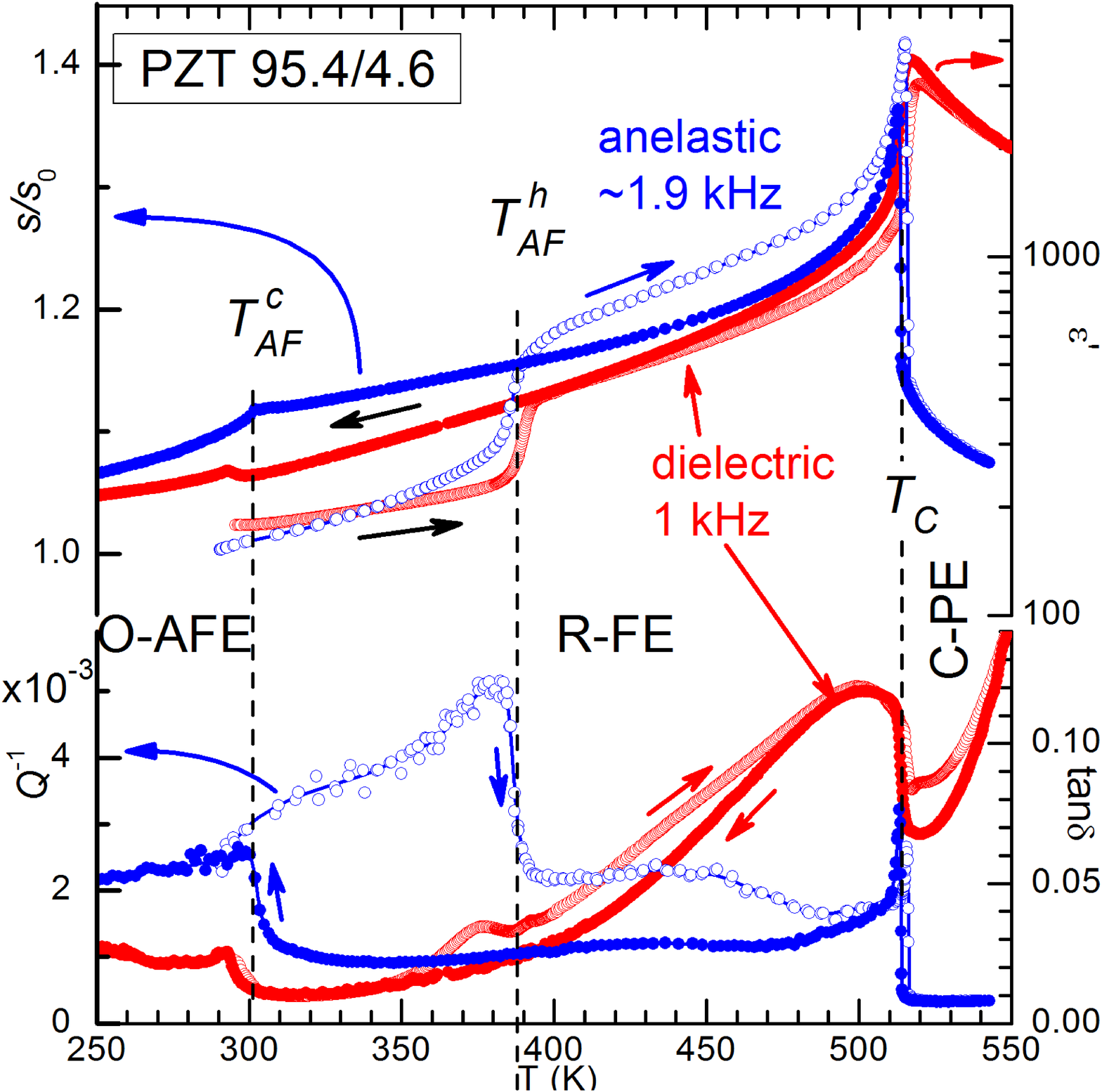}
\caption{Real parts (upper panel) and losses (lower panel) of the elastic
compliance and dielectric susceptibility of PZT 95.4/4.6.}
\label{fig_AnDiPZT}
\end{figure}

Figure \ref{fig_AnDiPZT} shows the real parts and losses of the dielectric
permittivity $\varepsilon $ (red) and elastic compliance $s$ (blue) of PZT
95.4/4.6, measured at 1~kHz and $\sim 1.9$~kHz respectively, during heating
(open symbols) and cooling (closed symbols). When cooling through $T_{%
\mathrm{C}}$ from the cubic (C) paraelectric (PE) to the rhombohedral (R) FE
phase, there is a drop of $\varepsilon ^{\prime }$ and positive jump of $%
s^{\prime }$. The jump in $\varepsilon ^{\prime }$ is due to the first order
nature of the transition, though the thermal hysteresis is only 2.5~K.
Notice the logarithmic scale of  $\varepsilon ^{\prime }$ in order to put in
evidence the anomalies at the AFE transitions.

The transition between the R-FE and the orthorhombic (O) AFE phase
has a large thermal hysteresis, and the susceptibility curves during
cooling may strongly depend on the cooling rate and state of the
sample \cite{CCT13,CTC14}. For this reason, the two temperatures
$T_{\mathrm{AF}}^{c}$ from the $s$ and $\varepsilon $ curves are
different, and the anomaly in $\varepsilon ^{\prime }$ has not the
same shape as during heating. Instead, the reverse
transition during heating occurs always at the same temperature $T_{\mathrm{%
AF}}^{h}$ accompanied by a steplike increase of the susceptibilities.

\begin{figure}[htb]
\includegraphics[width=8.5 cm]{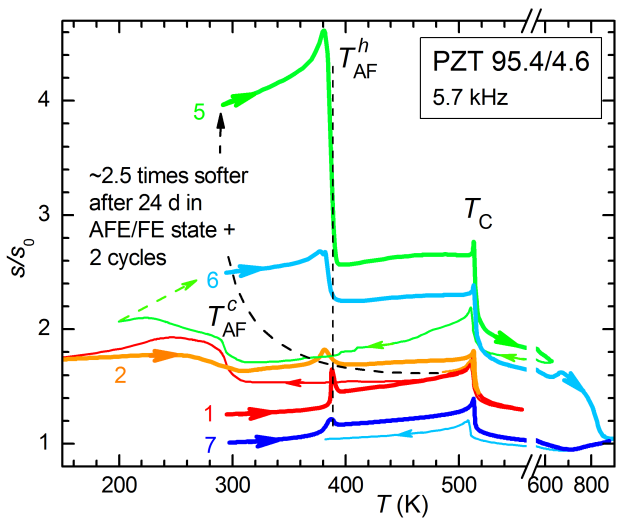}
\caption{Evolution of the compliance of PZT 95.4/4.6 during aging at room
temperature and the thermal cycles numbered in the labels.}
\label{fig_PZTcycl}
\end{figure}

Figure \ref{fig_PZTcycl} shows the evolution of the compliance of another
sample of PLZT 95.4/4.6 during thermal cycles and aging at room temperature
for overall one month. The sample of Fig. \ref{fig_AnDiPZT} also presents
similar behavior, but was subjected to a smaller number of cycles and for a
shorter time. The anomaly of $s^{\prime }$ at $T_{\mathrm{AF}}^{h}$ in Fig. %
\ref{fig_AnDiPZT} differs from those in Fig. \ref{fig_PZTcycl} for the lack
of a spike, which appeared in the fourth cycle (not shown). The shape of the
elastic anomaly at $T_{\mathrm{AF}}$ depends on the sample and its history
and can be fitted with the superposition of two steps of opposite signs
corresponding to the transitions in the polar and octahedral tilt modes;
apparently, the transition kinetics of these modes may be differently
affected by aging and fatigue, resulting in a variety of shapes \cite%
{CCT13,CTC14}. The spike arises from a slight shift in $T$ of the
centers of the two opposite steps. As described for similar
compositions around 5\% Ti \cite{CCT13,CTC14}, there is a
progressive softening associated both with aging at room
temperature, in a state of coexisting AFE and FE phases, and with
cycling through the temperature driven AFE/FE transitions. In the
sample of Fig. \ref{fig_PZTcycl} the compliance exhibits an
impressive softening of nearly four times at room temperature with
respect to the stiffer state just after a final high temperature
annealing. Not all the measured $s^{\prime }$ curves are displayed
for clarity; the numbers count the cycles that include runs through
$T_{\mathrm{AF}}^{h}$ and sometimes $T_{\mathrm{C}}$, without
exceeding it too much. In fact, when heated above 600~K, the sample
starts
recovering the original stiffness and shape of the anomaly at $T_{\mathrm{AF}%
}$, as shown by cycles 5 and 6. The latter was extended up to 850~K, causing
so complete an anneal, that the stiffness subsequently reached at room
temperature was even higher than at the starting point,
where some aging had already occurred. Notice
that no reoxygenation of excess O vacancies was possible during the high
temperature cycles, because they were made in a vacuum better than $10^{-5}$%
~mbar and at a fast rate, $5-6$~K/min above 600~K.

\begin{figure}[htb]
\includegraphics[width=8.5 cm]{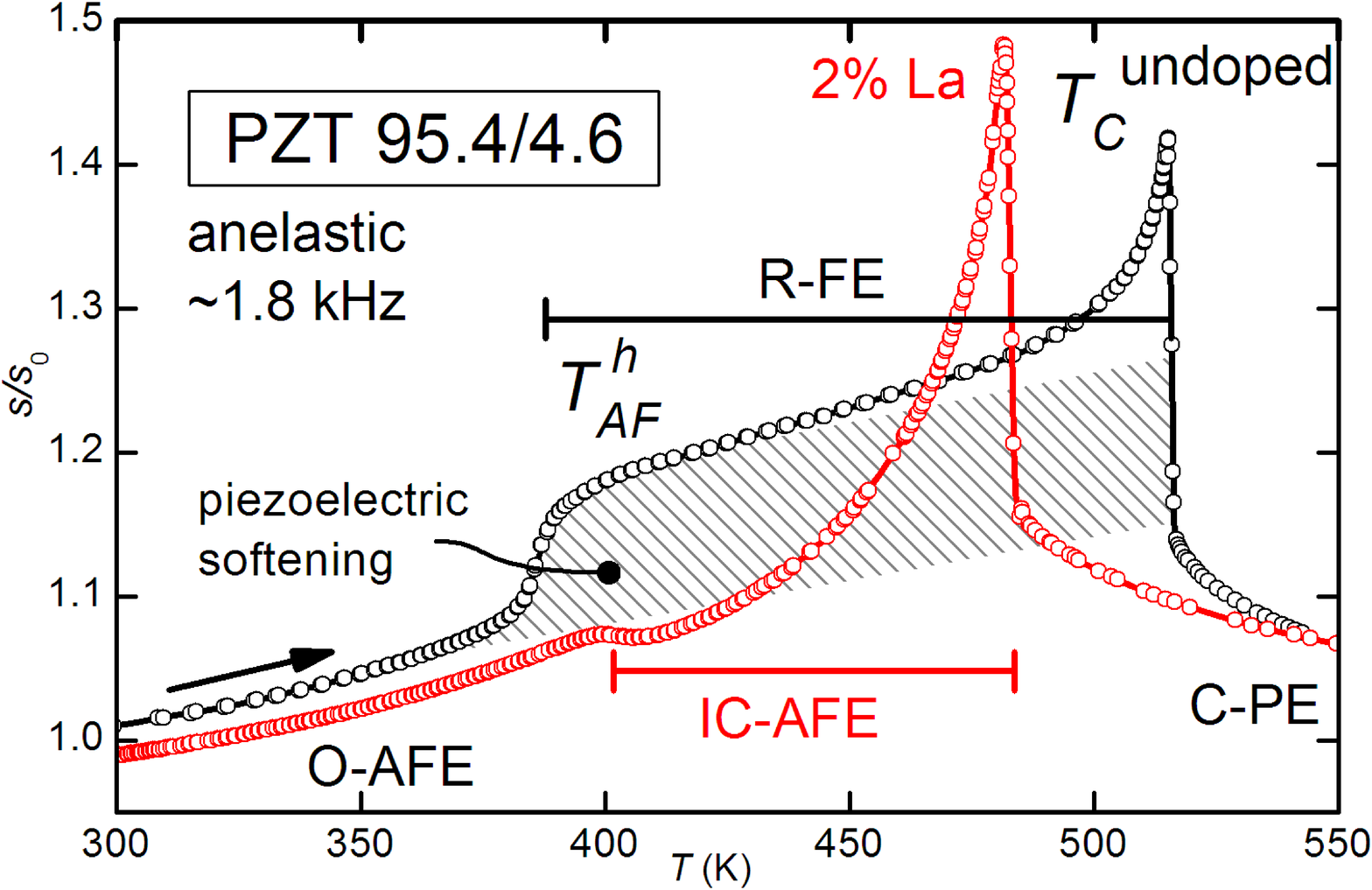}
\caption{Comparison of the compliances of PZT 95.4/4.6, undoped and doped
with 2\% La, measured during heating.}
\label{fig_AnPiezo}
\end{figure}

Similar tests have been conducted on PLZT 2/95.4/4.6. The
anelastic and dielectric curves in the virgin state are shown in Fig. \ref%
{fig_AnDiPLZT}.
\begin{figure}[htb]
\includegraphics[width=8.5 cm]{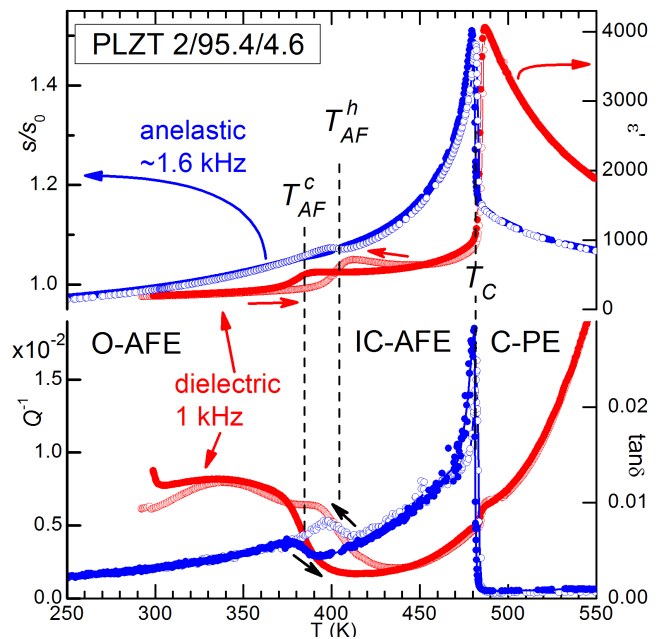}
\caption{Real parts (upper panel) and losses (lower panel) of the elastic
compliance and dielectric susceptibility of PLZT 2/95.4/4.6.}
\label{fig_AnDiPLZT}
\end{figure}
A major difference with the undoped composition is a closer
correspondence between the $\varepsilon $ curves during cooling and
heating, with reduced thermal hysteresis
$T_{\mathrm{AF}}^{h}-T_{\mathrm{AF}}^{c}$, indicating a faster
kinetics and no complications of splitting of antiferrodistortive
and polar modes below $T_{\mathrm{AF}}$. In addition, the elastic
response of the intermediate phase between $T_{\mathrm{AF}}$ and
$T_{\mathrm{C}}$ has changed, as put in evidence in Fig.
\ref{fig_AnPiezo}; here the $s^{\prime }$ heating curves of undoped
and La doped PZT are compared and the dashed region corresponds to
the piezoelectric softening \cite{CCT16} within the FE
phase of PZT, totally absent in PLZT, so that even the sign of the step at $%
T_{\mathrm{AF}}$ has changed.

\begin{figure}[htb]
\includegraphics[width=8.5 cm]{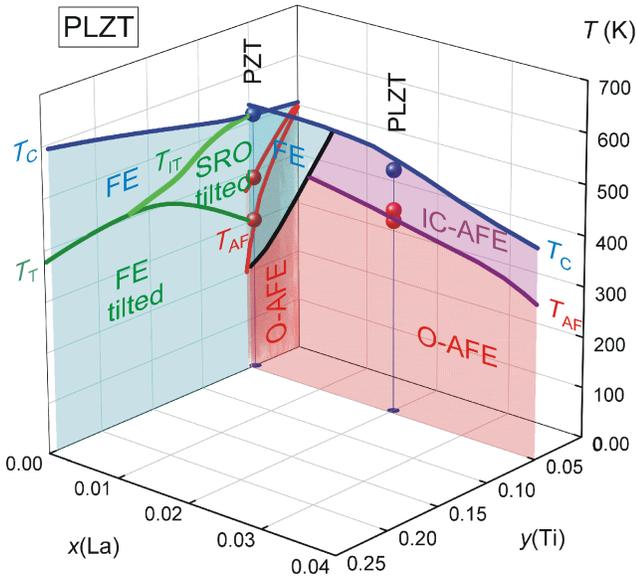}
\caption{Sections $x\left( \text{La}\right) =0$ \protect\cite{CTC13} and $%
y\left( \text{Ti}\right) =0.05$ \protect\cite{AK04} of the phase
diagram of PLZT. The points are the transitions temperatures of the
samples tested here. } \label{fig_PLZTpd}
\end{figure}

The disappearance of the piezoelectric softening with 2\% La doping
is in agreement with the PLZT $x/95/5$ phase diagram
\cite{DXV95b,AK04}, according to which for $x\left( \text{La}\right)
>0.01$ the intermediate R-FE phase
becomes IC-AFE. Sections of the $x\left( \text{La}\right) -y\left( \text{Ti}%
\right) -T$ phase diagram of PLZT are shown in Fig. \ref{fig_PLZTpd},
together with the transition temperatures of the present samples.

\begin{figure}[htb]
\includegraphics[width=8.5 cm]{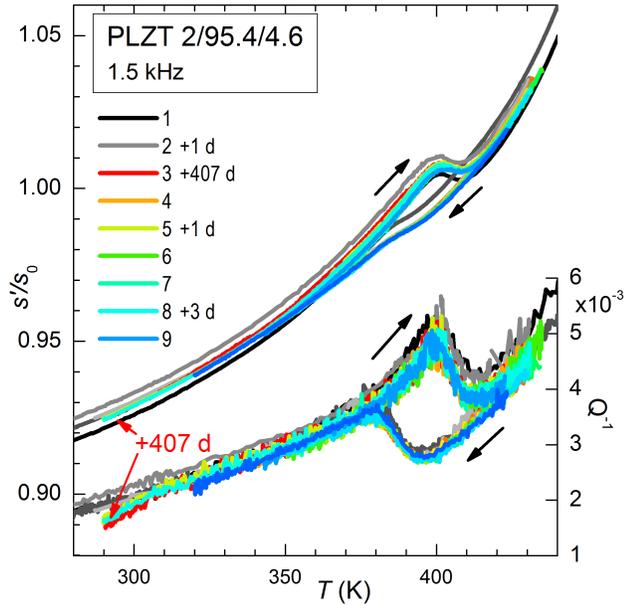}
\caption{Evolution of the compliance of PLZT 2/95.4/4.6 during aging at room
temperature in the AFE state and thermal cycles.}
\label{fig_PLZTcycl}
\end{figure}

The La-doped composition has a completely different behavior under aging at
room temperature and thermal cycling with respect to the undoped case, as
shown in Fig. \ref{fig_PLZTcycl}.

In contrast with PZT, PLZT remains remarkably constant over time and cycles:
the room temperature compliance remains within $\pm 0.3\%$ of the starting
value during 9 cycles, one of which extended to 760~K, distributed over 410
days, and the $s^{\prime }$ and $Q^{-1}$ steps at $T_{\mathrm{AF}}$ are
exactly reproducible. In addition, the amplitude of the $s^{\prime }$ step
is greatly reduced with respect to PZT.

\section{Discussion}

The ion La$^{3+}$ in coordination 12 has a radius of 136~pm, instead
of the 149~pm of Pb$^{2+}$, and the consequent reduction of the cell
volume is likely an important factor in the stabilization of the AFE
phase, that has a smaller volume than the FE one \cite{IS15}. This
is confirmed by the volumes of the pseudocubic cells deduced from
Rietveld refinements of the x-ray
diffractograms taken around $T_{\mathrm{AF}}$ and reported in Fig. \ref%
{fig_VvsT}. There is a more than tenfold decrease of the volume misfit between
the O-AFE and R-FE cells in PZT from $\left( 0.6 \pm 0.07\right) \%$ to
$\left( -0.04 \pm 0.06\right)\%$ for the O-AFE and IC-AFE cells in PLZT.

\begin{figure}[htb]
\includegraphics[width=8.5 cm]{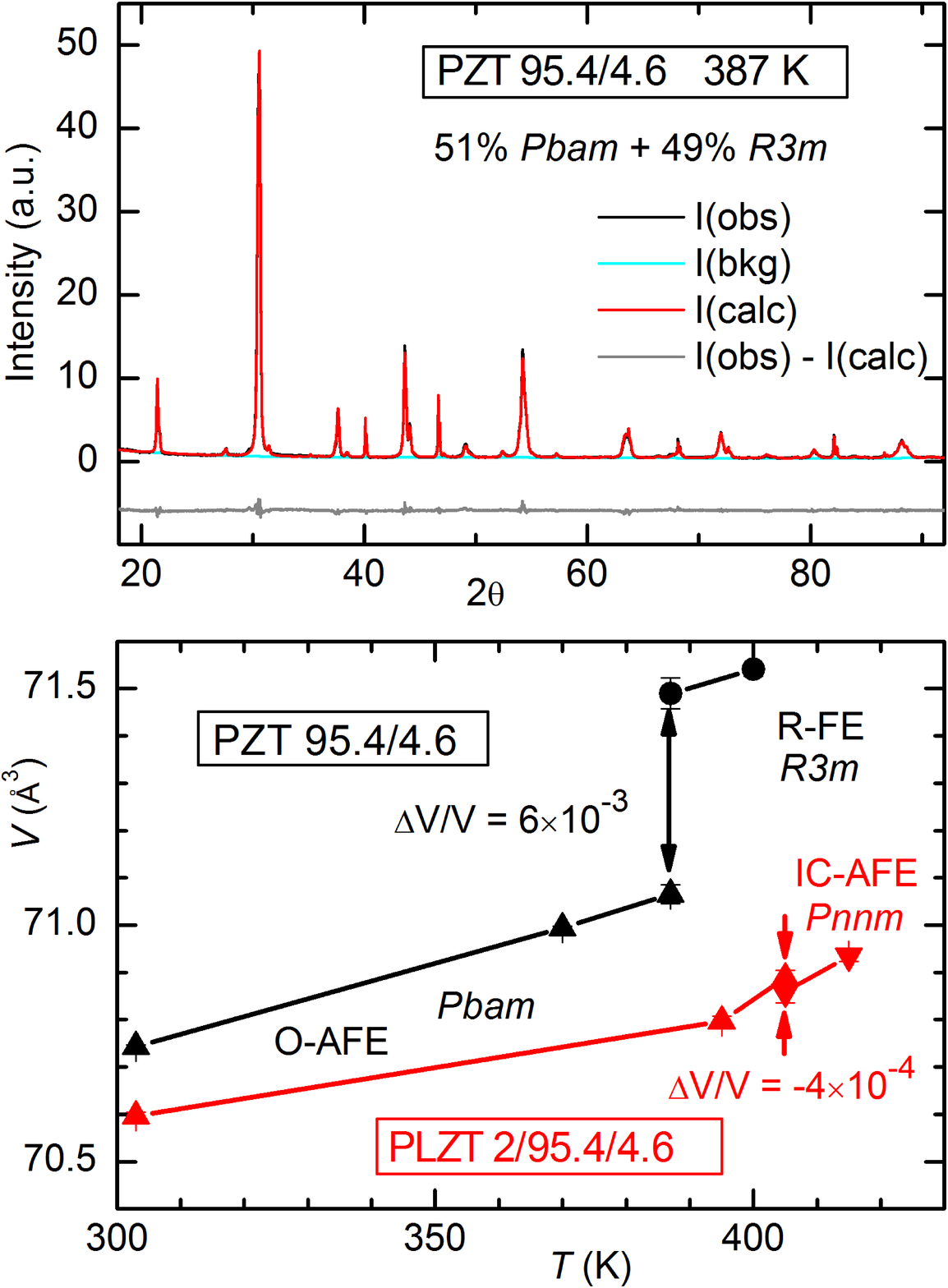}
\caption{Volumes of the pseudocubic unit cells R-FE ($R3m$), O-AFE ($Pbam$)
and IC-AFE ($Pnnm$) of PZT and PLZT around $T_{\mathrm{AF}}$. In the upper
panel is the refinement of the spectrum of PZT at $T_{\mathrm{AF}}$ with
51\% \textit{Pbam\ }+ 49\% \textit{R3m}.}
\label{fig_VvsT}
\end{figure}

In addition, La$^{3+}$ on the Pb$^{2+}$ site, accompanied by Pb
vacancies for charge compensation, is thought to destabilize the
long range FE\ order \cite{DV94}, because it introduces charge
disorder and lacks the lone-pair that drives Pb$^{2+}$ off-center.
In the phase diagram of Asada and Koyama \cite{AK04}, which
constitutes the $\left( x=0.05\right) -y-T$ section in Fig.
\ref{fig_PLZTpd}, the FE phase is found to be incommensurate, due to
the concurrence of the AFE modes of Pb and of octahedral tilting.
Note that also in La-free PZT with $y\left( \text{Ti}\right) <0.15$
the R-FE phase undergoes a phase transition below $T_{\mathrm{IT}}$,
which is evident in the compliance and therefore identified with
non-polar short-range ordered (SRO) tilting of the octahedra
\cite{CTC13}. The volume effect and the FE
destabilization result in a rise of $T_{\mathrm{AF}}$ with $x\left( \text{La}%
\right) $ and change of the intermediate phase from R-FE to IC-AFE.

That the nature of the intermediate phase below $T_{\mathrm{C}}$ is changed
by 2\% La doping is evident in the associated elastic anomaly (Fig. \ref%
{fig_AnPiezo}). In undoped PZT, when cooling from the PE phase one
finds a very extended precursor softening ending in a spike at
$T_{\mathrm{C}}$ and the compliance remains larger (or the modulus
softer) in the FE phase. The softening in the FE\ phase is of
piezoelectric origin \cite{CCT16}, and is
therefore absent in the AFE phase. In fact, in Fig. \ref{fig_AnPiezo} the $%
s^{\prime }$ curves of PZT and PLZT are very similar to each other, except
for the lack of a higher plateau in the intermediate phase of PLZT.

More striking is the difference in the behaviors of the compliance during
aging for the two compositions, illustrated in Figs. \ref{fig_PZTcycl} and %
\ref{fig_PLZTcycl}. The compliance of PZT at room temperature spans a range
of nearly 400\% of the stiffer value in the virgin state, exceptional for a
ceramic, while PLZT remains stable within $\pm 0.3\%$ after passing a larger
number of cycles and 14 times longer aging.

The comparison between the two samples demonstrates that the elastic
aging and fatigue in undoped PZT, the latter realized by the
thermally induced transitions, are not simply related to the O-AFE
phase, which is common to the two compositions, but are due to its
coexistence with the FE phase. As a consequence, aging and fatigue
must be connected to defects forming at the FE/AFE interfaces. The
difference in volume between AFE and FE phases is the core of the
explanations proposed for various aging and irreversible effects
observed in diffraction and dielectric measurements of PZT-based
materials \cite{PP99,IS15}. The accommodation of the misfit strain
has been proposed to cause broken and dangling bonds \cite{PP99} and
to promote the migration of the larger Zr$^{4+}$cations to the AFE
domains and smaller Ti$^{4+}$ and La$^{3+}$ cations to the FE
domains \cite{IS15}. Also microcracking can be found in fatigued FE
and AFE ceramics \cite{JSC94,ZZZ04}, but neither microcracking nor a
large/small cation imbalance should be the cause of the elastic
aging discussed here, because they are not expected to be recovered
by mild annealing well below the sintering temperature. In fact, the
strong electric and elastic fields that might possibly promote the
cation migration during the coexistence of the AFE and FE domains
are missing in the C-PE phase, and under such conditions the cation
migration occurs only at the sintering temperature. Also the similar
observation of partial recovery of electrically fatigued PLZST after
heating at $\sim 770$~K for 1~h has been discussed in terms of
migration of V$_{\text{O}}$ or other charged species, rather than
microcrack healing \cite{ZZZ04}.

The total cancelation of elastic aging and fatigue with 2\% La
doping is likely the result of multiple mechanisms. First of all,
the FE phase becomes IC-AFE, whose smaller volume is closer to that
of the O-AFE phase, and therefore reduces the misfit between the
coexisting phases when passing through the transition and aging
within its region of thermal hysteresis. In addition to relieving
the mechanical stresses, the substitution of the R-FE phase with
IC-AFE, eliminates the electric fields at the borders of the FE
domains, where the polarization charge is distributed \cite{Gen08}.
Both the
stress and electric fields would drive mobile defects like O vacancies \cite%
{GL07} (V$_{\text{O}}$), which posses different charge and specific volume
with respect to O$^{2-}$. Though we do not have a quantitative model, we
think that the main mechanism behind the elastic aging and fatigue effects
is the migration of V$_{\text{O}}$ to the charged and stressed walls between
AFE and FE phases, followed by aggregation into relatively stable structures.%
\emph{\ }Concentrations of few tenths of percent of V$_{\text{O}}$ are
always present as a result of the PbO loss during sintering, and their
hypothetical linear or planar aggregations would certainly dissolve during
moderate annealing. Note that the annealings in Fig. \ref{fig_PZTcycl} are
made in a vacuum of $<10^{-5}$~mbar, and therefore cannot fill V$_{\text{O}}$%
, but simply dissolve their aggregations. It is not obvious,
however, how aggregations of small concentrations of V$_{\text{O}}$,
likely well below 1\%, would be able to soften the overall Young's
modulus as much as observed. This remains an important point to be
investigated, possibly with TEM and related techniques able to
detect defect structures at the atomic level \cite{WRY04}.

As an additional mechanism, that contributes to preventing aging and
fatigue in the La-doped material, one should consider the reduction
or neutralization of the mobile defects. In fact, La$^{3+}$
substituting Pb$^{2+}$ is a donor, and makes the material
electrically soft by improving the domain wall mobility, a
phenomenon not well understood \cite{Dam98} but implicitly
attributed to a reduction of the concentration of V$_{\text{O}}$
that pin the domains.

\begin{figure}[htb]
\includegraphics[width=8.5 cm]{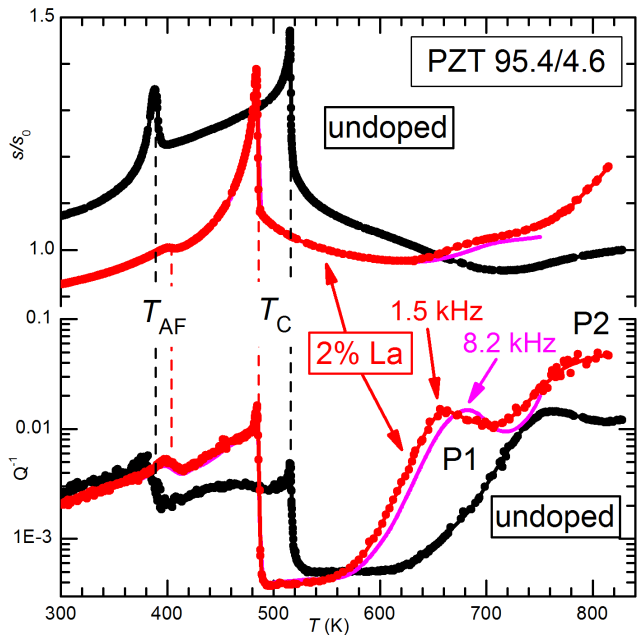}
\caption{Anelastic spectra of undoped and 2\% La doped PZT 95.4/4.6 extended
to high temperature. Of the doped sample are shown the curves measured
during the same run at $\sim 1.5$ and $\sim 8.2$~kHz; the curves coincide
except for the thermally activated relaxation processes P1 and P2.}
\label{fig_anel13HT}
\end{figure}

In principle, the hopping of V$_{\text{O}}$ should produce anelastic
relaxation with one or more relaxation rates $\tau ^{-1}$, detectable as one
or more peaks in the $Q^{-1}$ at the temperatures $T_{m}$ and frequencies $%
\omega /2\pi $ satisfying $\omega \tau \left( T_{m}\right) \simeq 1$ \cite%
{NB72}, as found in SrTiO$_{3}$ \cite{130}. Yet, already the
apparently simple case of cubic SrTiO$_{3}$ presents a complicated
anelastic spectrum, due to isolated and variously aggregated
V$_{\text{O}}$, with activation energies spanning from 0.6 to 1~eV;
PLZT has the additional complication of strong traps, like Pb
vacancies, and other perturbations like La$^{3+}$ substituting
Pb$^{2+}$. Moreover, $Q^{-1}$ maxima from relaxation processes with
activation energies $E<0.9$~eV are expected to appear at
$T_{m}<500$~K, assuming $\tau =\tau _{0}\exp \left( E/T\right) $
with $\tau _{0}\sim 10^{-13}$~s and $f\sim 2$~kHz, and would be
masked by the losses associated with the domain wall motion below $T_{%
\mathrm{C}}$. Indeed, we always observe at least two thermally activated $%
Q^{-1}\left( T\right) $ peaks in the cubic phase of Zr-rich PZT, indicated
as P1 and P2 in Fig. \ref{fig_anel13HT}, and at least P1 at lower
temperature may be due to hopping of V$_{\text{O}}$, though not simply isolated
V$_{\text{O}}$ in a regular lattice as in SrTiO$_{3}$. In fact, the high
apparent activation energy, deduced from the shift of the peak in temperature
when frequency changes, indicates correlated motions, similarly to
the processes described by the Vogel-Fulcher law $\tau =\tau _{0}\exp
E/\left( T-T_{0}\right) $, where the slope of $\ln \tau $ vs $1/T$ exceeds $E
$ on approaching $T_{0}$. The high temperature anelastic spectra of the two
samples are shown in Fig. \ref{fig_anel13HT} and, without attempting
interpretations of the nature of the thermally activated peaks P1 and P2, we
can say that they are due to mobile defects and P1 is most likely due to V$_{%
\text{O}}$, because the cations should not be able to perform $10^{4}$ jumps
per second below 700~K. Lanthanum doping does not suppress the heights of
these peaks, but only shifts them to lower temperature, implying that about
the same concentrations of defects is present, but with higher mobility.
This observation seems to rule out a suppression of V$_{\text{O}}$ by La
doping, at least in this case.

\section{Conclusions}

It is usually found that the AFE materials withstand repetitive
electric cycling better than the FE materials \cite{JSC94,ZZR06}
with some compositions withstanding $10^{7}$ cycles before
appreciable decrease in the saturation polarization and strain. Yet,
the Young's modulus of AFE PZT may suffer huge softenings, up to a
factor of $1/4$, simply upon aging at room temperature few weeks
with few thermally induced AFE/FE transitions. Though the
temperature induced AFE/FE transition are different from the field
induced ones, which are thought to involve only strain-free
180$^{\circ }$ switching, the magnitude of the presently reported
phenomena suggests that materials for applications based on AFE/FE
transitions, especially when induced by temperature changes, should
be elastically tested also over long aging periods.

Doping with 2\% La transforms the intermediate FE phase into IC-AFE, and
completely suppresses the elastic aging: the Young's modulus is stable
within $\pm 0.3\%$ over more than one year and several thermal cycles.

It is proposed that the migration of O vacancies under the electric and stress
fields at the AFE/FE interfaces and their aggregation into stable structures
play a role in the elastic aging, since rapid annealing in vacuum at 800~K
is sufficient to completely recover the original stiffness.

The awareness of so large a dependence of the elastic moduli on the sample
history and of the presence of recoverable extended defects should be
important when studying the methods for limiting the crack formation and
propagation in these materials \cite{Uch98,TYS14}.

\begin{acknowledgments}
The authors thank C. Capiani (ISTEC) for the preparation of the
samples, Chiara Zanelli (ISTEC) for collecting the XRD spectra, and
P. M. Latino (ISC) for his valuable technical assistance.
\end{acknowledgments}

\bibliography{refs}

\begin{thebibliography}{35}%
\makeatletter
\providecommand \@ifxundefined [1]{%
 \@ifx{#1\undefined}
}%
\providecommand \@ifnum [1]{%
 \ifnum #1\expandafter \@firstoftwo
 \else \expandafter \@secondoftwo
 \fi
}%
\providecommand \@ifx [1]{%
 \ifx #1\expandafter \@firstoftwo
 \else \expandafter \@secondoftwo
 \fi
}%
\providecommand \natexlab [1]{#1}%
\providecommand \enquote  [1]{``#1''}%
\providecommand \bibnamefont  [1]{#1}%
\providecommand \bibfnamefont [1]{#1}%
\providecommand \citenamefont [1]{#1}%
\providecommand \href@noop [0]{\@secondoftwo}%
\providecommand \href [0]{\begingroup \@sanitize@url \@href}%
\providecommand \@href[1]{\@@startlink{#1}\@@href}%
\providecommand \@@href[1]{\endgroup#1\@@endlink}%
\providecommand \@sanitize@url [0]{\catcode `\\12\catcode `\$12\catcode
  `\&12\catcode `\#12\catcode `\^12\catcode `\_12\catcode `\%12\relax}%
\providecommand \@@startlink[1]{}%
\providecommand \@@endlink[0]{}%
\providecommand \url  [0]{\begingroup\@sanitize@url \@url }%
\providecommand \@url [1]{\endgroup\@href {#1}{\urlprefix }}%
\providecommand \urlprefix  [0]{URL }%
\providecommand \Eprint [0]{\href }%
\providecommand \doibase [0]{http://dx.doi.org/}%
\providecommand \selectlanguage [0]{\@gobble}%
\providecommand \bibinfo  [0]{\@secondoftwo}%
\providecommand \bibfield  [0]{\@secondoftwo}%
\providecommand \translation [1]{[#1]}%
\providecommand \BibitemOpen [0]{}%
\providecommand \bibitemStop [0]{}%
\providecommand \bibitemNoStop [0]{.\EOS\space}%
\providecommand \EOS [0]{\spacefactor3000\relax}%
\providecommand \BibitemShut  [1]{\csname bibitem#1\endcsname}%
\let\auto@bib@innerbib\@empty
%</preamble>
\bibitem [{\citenamefont {Hao}\ \emph {et~al.}(2014)\citenamefont {Hao},
  \citenamefont {Zhai}, \citenamefont {Kong},\ and\ \citenamefont
  {Xu}}]{HZK14}%
  \BibitemOpen
  \bibfield  {author} {\bibinfo {author} {\bibfnamefont {X.}~\bibnamefont
  {Hao}}, \bibinfo {author} {\bibfnamefont {J.}~\bibnamefont {Zhai}}, \bibinfo
  {author} {\bibfnamefont {L.~B.}\ \bibnamefont {Kong}}, \ and\ \bibinfo
  {author} {\bibfnamefont {Z.}~\bibnamefont {Xu}},\ }\href@noop {} {\bibfield
  {journal} {\bibinfo  {journal} {Progr. Mater. Sci.}\ }\textbf {\bibinfo
  {volume} {63}},\ \bibinfo {pages} {1} (\bibinfo {year} {2014})}\BibitemShut
  {NoStop}%
\bibitem [{\citenamefont {Tan}\ \emph {et~al.}(2011)\citenamefont {Tan},
  \citenamefont {Ma}, \citenamefont {Frederick}, \citenamefont {Beckman},\ and\
  \citenamefont {Webber}}]{TMF11}%
  \BibitemOpen
  \bibfield  {author} {\bibinfo {author} {\bibfnamefont {X.}~\bibnamefont
  {Tan}}, \bibinfo {author} {\bibfnamefont {C.}~\bibnamefont {Ma}}, \bibinfo
  {author} {\bibfnamefont {J.}~\bibnamefont {Frederick}}, \bibinfo {author}
  {\bibfnamefont {S.}~\bibnamefont {Beckman}}, \ and\ \bibinfo {author}
  {\bibfnamefont {K.~G.}\ \bibnamefont {Webber}},\ }\href@noop {} {\bibfield
  {journal} {\bibinfo  {journal} {J. Am. Ceram. Soc.}\ }\textbf {\bibinfo
  {volume} {94}},\ \bibinfo {pages} {4091} (\bibinfo {year}
  {2011})}\BibitemShut {NoStop}%
\bibitem [{\citenamefont {Hao}\ \emph {et~al.}(2015)\citenamefont {Hao},
  \citenamefont {Zhao},\ and\ \citenamefont {Zhang}}]{HZZ15}%
  \BibitemOpen
  \bibfield  {author} {\bibinfo {author} {\bibfnamefont {X.}~\bibnamefont
  {Hao}}, \bibinfo {author} {\bibfnamefont {Y.}~\bibnamefont {Zhao}}, \ and\
  \bibinfo {author} {\bibfnamefont {Q.}~\bibnamefont {Zhang}},\ }\href@noop {}
  {\bibfield  {journal} {\bibinfo  {journal} {J. Phys. Chem. C}\ }\textbf
  {\bibinfo {volume} {119}},\ \bibinfo {pages} {18877} (\bibinfo {year}
  {2015})}\BibitemShut {NoStop}%
\bibitem [{\citenamefont {Mischenko}\ \emph {et~al.}(2006)\citenamefont
  {Mischenko}, \citenamefont {Zhang}, \citenamefont {Scott}, \citenamefont
  {Whatmore},\ and\ \citenamefont {Mathur}}]{MZS06}%
  \BibitemOpen
  \bibfield  {author} {\bibinfo {author} {\bibfnamefont {A.~S.}\ \bibnamefont
  {Mischenko}}, \bibinfo {author} {\bibfnamefont {Q.}~\bibnamefont {Zhang}},
  \bibinfo {author} {\bibfnamefont {J.~F.}\ \bibnamefont {Scott}}, \bibinfo
  {author} {\bibfnamefont {R.~W.}\ \bibnamefont {Whatmore}}, \ and\ \bibinfo
  {author} {\bibfnamefont {N.~D.}\ \bibnamefont {Mathur}},\ }\href@noop {}
  {\bibfield  {journal} {\bibinfo  {journal} {Science}\ }\textbf {\bibinfo
  {volume} {311}},\ \bibinfo {pages} {1270} (\bibinfo {year}
  {2006})}\BibitemShut {NoStop}%
\bibitem [{\citenamefont {Hao}\ \emph {et~al.}(2011)\citenamefont {Hao},
  \citenamefont {Yue}, \citenamefont {Xu}, \citenamefont {An},\ and\
  \citenamefont {Nan}}]{HYX11}%
  \BibitemOpen
  \bibfield  {author} {\bibinfo {author} {\bibfnamefont {X.}~\bibnamefont
  {Hao}}, \bibinfo {author} {\bibfnamefont {Z.}~\bibnamefont {Yue}}, \bibinfo
  {author} {\bibfnamefont {J.}~\bibnamefont {Xu}}, \bibinfo {author}
  {\bibfnamefont {S.}~\bibnamefont {An}}, \ and\ \bibinfo {author}
  {\bibfnamefont {C.~W.}\ \bibnamefont {Nan}},\ }\href@noop {} {\bibfield
  {journal} {\bibinfo  {journal} {J. Appl. Phys.}\ }\textbf {\bibinfo {volume}
  {110}},\ \bibinfo {pages} {064109} (\bibinfo {year} {2011})}\BibitemShut
  {NoStop}%
\bibitem [{\citenamefont {F{\"a}hler}\ \emph {et~al.}(2012)\citenamefont
  {F{\"a}hler}, \citenamefont {R{\"o}{\ss}ler}, \citenamefont {Kastner},
  \citenamefont {Eckert}, \citenamefont {Eggeler}, \citenamefont {Emmerich},
  \citenamefont {Entel}, \citenamefont {M{\"u}ller}, \citenamefont {Quandt},\
  and\ \citenamefont {Albe}}]{FRK12}%
  \BibitemOpen
  \bibfield  {author} {\bibinfo {author} {\bibfnamefont {S.}~\bibnamefont
  {F{\"a}hler}}, \bibinfo {author} {\bibfnamefont {U.~K.}\ \bibnamefont
  {R{\"o}{\ss}ler}}, \bibinfo {author} {\bibfnamefont {O.}~\bibnamefont
  {Kastner}}, \bibinfo {author} {\bibfnamefont {J.}~\bibnamefont {Eckert}},
  \bibinfo {author} {\bibfnamefont {G.}~\bibnamefont {Eggeler}}, \bibinfo
  {author} {\bibfnamefont {H.}~\bibnamefont {Emmerich}}, \bibinfo {author}
  {\bibfnamefont {P.}~\bibnamefont {Entel}}, \bibinfo {author} {\bibfnamefont
  {S.}~\bibnamefont {M{\"u}ller}}, \bibinfo {author} {\bibfnamefont
  {E.}~\bibnamefont {Quandt}}, \ and\ \bibinfo {author} {\bibfnamefont
  {K.}~\bibnamefont {Albe}},\ }\href@noop {} {\bibfield  {journal} {\bibinfo
  {journal} {Adv. Eng. Mater.}\ }\textbf {\bibinfo {volume} {14}},\ \bibinfo
  {pages} {10} (\bibinfo {year} {2012})}\BibitemShut {NoStop}%
\bibitem [{\citenamefont {Takeuchi}\ and\ \citenamefont
  {Sandeman}(2015)}]{TS15}%
  \BibitemOpen
  \bibfield  {author} {\bibinfo {author} {\bibfnamefont {I.}~\bibnamefont
  {Takeuchi}}\ and\ \bibinfo {author} {\bibfnamefont {K.}~\bibnamefont
  {Sandeman}},\ }\href@noop {} {\bibfield  {journal} {\bibinfo  {journal}
  {Physics Today}\ }\textbf {\bibinfo {volume} {68}},\ \bibinfo {pages} {48}
  (\bibinfo {year} {2015})}\BibitemShut {NoStop}%
\bibitem [{\citenamefont {Moya}\ \emph {et~al.}(2014)\citenamefont {Moya},
  \citenamefont {Kar-Narayan},\ and\ \citenamefont {Mathur}}]{MKM14}%
  \BibitemOpen
  \bibfield  {author} {\bibinfo {author} {\bibfnamefont {X.}~\bibnamefont
  {Moya}}, \bibinfo {author} {\bibfnamefont {S.}~\bibnamefont {Kar-Narayan}}, \
  and\ \bibinfo {author} {\bibfnamefont {N.~D.}\ \bibnamefont {Mathur}},\
  }\href@noop {} {\bibfield  {journal} {\bibinfo  {journal} {Nat. Mater.}\
  }\textbf {\bibinfo {volume} {13}},\ \bibinfo {pages} {439} (\bibinfo {year}
  {2014})}\BibitemShut {NoStop}%
\bibitem [{\citenamefont {Pokharel}\ and\ \citenamefont {Pandey}(1999)}]{PP99}%
  \BibitemOpen
  \bibfield  {author} {\bibinfo {author} {\bibfnamefont {B.~P.}\ \bibnamefont
  {Pokharel}}\ and\ \bibinfo {author} {\bibfnamefont {D.}~\bibnamefont
  {Pandey}},\ }\href@noop {} {\bibfield  {journal} {\bibinfo  {journal} {J.
  Appl. Phys.}\ }\textbf {\bibinfo {volume} {86}},\ \bibinfo {pages} {3327}
  (\bibinfo {year} {1999})}\BibitemShut {NoStop}%
\bibitem [{\citenamefont {Cordero}\ \emph
  {et~al.}(2013{\natexlab{a}})\citenamefont {Cordero}, \citenamefont {Craciun},
  \citenamefont {Trequattrini}, \citenamefont {Galassi}, \citenamefont
  {Thomas}, \citenamefont {Keeble},\ and\ \citenamefont {Glazer}}]{CCT13}%
  \BibitemOpen
  \bibfield  {author} {\bibinfo {author} {\bibfnamefont {F.}~\bibnamefont
  {Cordero}}, \bibinfo {author} {\bibfnamefont {F.}~\bibnamefont {Craciun}},
  \bibinfo {author} {\bibfnamefont {F.}~\bibnamefont {Trequattrini}}, \bibinfo
  {author} {\bibfnamefont {C.}~\bibnamefont {Galassi}}, \bibinfo {author}
  {\bibfnamefont {P.~A.}\ \bibnamefont {Thomas}}, \bibinfo {author}
  {\bibfnamefont {D.~S.}\ \bibnamefont {Keeble}}, \ and\ \bibinfo {author}
  {\bibfnamefont {A.~M.}\ \bibnamefont {Glazer}},\ }\href@noop {} {\bibfield
  {journal} {\bibinfo  {journal} {Phys. Rev. B}\ }\textbf {\bibinfo {volume}
  {88}},\ \bibinfo {pages} {094107} (\bibinfo {year}
  {2013}{\natexlab{a}})}\BibitemShut {NoStop}%
\bibitem [{\citenamefont {Cordero}\ \emph {et~al.}(2014)\citenamefont
  {Cordero}, \citenamefont {Trequattrini}, \citenamefont {Craciun},\ and\
  \citenamefont {Galassi}}]{CTC14}%
  \BibitemOpen
  \bibfield  {author} {\bibinfo {author} {\bibfnamefont {F.}~\bibnamefont
  {Cordero}}, \bibinfo {author} {\bibfnamefont {F.}~\bibnamefont
  {Trequattrini}}, \bibinfo {author} {\bibfnamefont {F.}~\bibnamefont
  {Craciun}}, \ and\ \bibinfo {author} {\bibfnamefont {C.}~\bibnamefont
  {Galassi}},\ }\href@noop {} {\bibfield  {journal} {\bibinfo  {journal} {Phys.
  Rev. B}\ }\textbf {\bibinfo {volume} {89}},\ \bibinfo {pages} {214102}
  (\bibinfo {year} {2014})}\BibitemShut {NoStop}%
\bibitem [{Note1()}]{Note1}%
  \BibitemOpen
  \bibinfo {note} {In Refs. \protect \rev@citealpnum {CCT13,CTC14} we
  identified the stiffening on cooling with the orthorhombic tilt mode (OT) and
  the softening with the antiferroelectric mode (AF). At that time, as only
  explanation for a steplike stiffening during cooling through a structural
  phase transition we found the loss of one tilt system from the disordered
  tilted phase (with octahedral rotations about all axes) to the orthorhombic
  phase, untilted along the $c$ axis. Now we are convinced of the importance of
  the piezoelectric softening in the FE phase, and therefore attribute the
  stiffening on cooling to the AF mode and the softening to the OT mode. What
  is written in Refs. \protect \rev@citealpnum {CCT13,CTC14} is valid if one
  exchanges AF with OT everywhere, except for some inessential comments to few
  fittings. Some of the fits of the $s^{\prime }\left ( T\right ) $ curves are
  slightly different, due to the compatibility conditions between the different
  phases embedded in the fitting equations, but the parameters remain identical
  or very similar with AF and OT exchanged. Also the discussion of the effect
  of the loss of tilting about the $c$ axis remains valid, though evidently it
  is not enough to transform into stiffening the expected softening at the
  onset of the long range order below $T_{\protect \mathrm {OT}}$.}\BibitemShut
  {Stop}%
\bibitem [{\citenamefont {Dai}\ \emph {et~al.}(1995)\citenamefont {Dai},
  \citenamefont {Xu},\ and\ \citenamefont {Viehland}}]{DXV95b}%
  \BibitemOpen
  \bibfield  {author} {\bibinfo {author} {\bibfnamefont {X.}~\bibnamefont
  {Dai}}, \bibinfo {author} {\bibfnamefont {Z.}~\bibnamefont {Xu}}, \ and\
  \bibinfo {author} {\bibfnamefont {D.}~\bibnamefont {Viehland}},\ }\href@noop
  {} {\bibfield  {journal} {\bibinfo  {journal} {J. Appl. Phys.}\ }\textbf
  {\bibinfo {volume} {77}},\ \bibinfo {pages} {5086} (\bibinfo {year}
  {1995})}\BibitemShut {NoStop}%
\bibitem [{\citenamefont {Asada}\ and\ \citenamefont {Koyama}(2004)}]{AK04}%
  \BibitemOpen
  \bibfield  {author} {\bibinfo {author} {\bibfnamefont {T.}~\bibnamefont
  {Asada}}\ and\ \bibinfo {author} {\bibfnamefont {Y.}~\bibnamefont {Koyama}},\
  }\href@noop {} {\bibfield  {journal} {\bibinfo  {journal} {Phys. Rev. B}\
  }\textbf {\bibinfo {volume} {69}},\ \bibinfo {pages} {104108} (\bibinfo
  {year} {2004})}\BibitemShut {NoStop}%
\bibitem [{\citenamefont {Cordero}\ \emph {et~al.}(2007)\citenamefont
  {Cordero}, \citenamefont {Craciun},\ and\ \citenamefont {Galassi}}]{127}%
  \BibitemOpen
  \bibfield  {author} {\bibinfo {author} {\bibfnamefont {F.}~\bibnamefont
  {Cordero}}, \bibinfo {author} {\bibfnamefont {F.}~\bibnamefont {Craciun}}, \
  and\ \bibinfo {author} {\bibfnamefont {C.}~\bibnamefont {Galassi}},\
  }\href@noop {} {\bibfield  {journal} {\bibinfo  {journal} {Phys. Rev. Lett.}\
  }\textbf {\bibinfo {volume} {98}},\ \bibinfo {pages} {255701} (\bibinfo
  {year} {2007})}\BibitemShut {NoStop}%
\bibitem [{\citenamefont {Cordero}\ \emph {et~al.}(2011)\citenamefont
  {Cordero}, \citenamefont {Trequattrini}, \citenamefont {Craciun},\ and\
  \citenamefont {Galassi}}]{145}%
  \BibitemOpen
  \bibfield  {author} {\bibinfo {author} {\bibfnamefont {F.}~\bibnamefont
  {Cordero}}, \bibinfo {author} {\bibfnamefont {F.}~\bibnamefont
  {Trequattrini}}, \bibinfo {author} {\bibfnamefont {F.}~\bibnamefont
  {Craciun}}, \ and\ \bibinfo {author} {\bibfnamefont {C.}~\bibnamefont
  {Galassi}},\ }\href@noop {} {\bibfield  {journal} {\bibinfo  {journal} {J.
  Phys.: Condens. Matter}\ }\textbf {\bibinfo {volume} {23}},\ \bibinfo {pages}
  {415901} (\bibinfo {year} {2011})}\BibitemShut {NoStop}%
\bibitem [{\citenamefont {Corker}\ \emph {et~al.}(1997)\citenamefont {Corker},
  \citenamefont {Glazer}, \citenamefont {Dec}, \citenamefont {Roleder},\ and\
  \citenamefont {Whatmore}}]{CGD97}%
  \BibitemOpen
  \bibfield  {author} {\bibinfo {author} {\bibfnamefont {D.~L.}\ \bibnamefont
  {Corker}}, \bibinfo {author} {\bibfnamefont {A.~M.}\ \bibnamefont {Glazer}},
  \bibinfo {author} {\bibfnamefont {J.}~\bibnamefont {Dec}}, \bibinfo {author}
  {\bibfnamefont {K.}~\bibnamefont {Roleder}}, \ and\ \bibinfo {author}
  {\bibfnamefont {R.~W.}\ \bibnamefont {Whatmore}},\ }\href@noop {} {\bibfield
  {journal} {\bibinfo  {journal} {Acta Cryst. B}\ }\textbf {\bibinfo {volume}
  {53}},\ \bibinfo {pages} {135} (\bibinfo {year} {1997})}\BibitemShut
  {NoStop}%
\bibitem [{\citenamefont {Liu}\ and\ \citenamefont {Toraya}(1999)}]{LT99b}%
  \BibitemOpen
  \bibfield  {author} {\bibinfo {author} {\bibfnamefont {H.}~\bibnamefont
  {Liu}}\ and\ \bibinfo {author} {\bibfnamefont {H.}~\bibnamefont {Toraya}},\
  }\href@noop {} {\bibfield  {journal} {\bibinfo  {journal} {Jpn. J. Appl.
  Phys.}\ }\textbf {\bibinfo {volume} {38}},\ \bibinfo {pages} {104} (\bibinfo
  {year} {1999})}\BibitemShut {NoStop}%
\bibitem [{\citenamefont {Michel}\ \emph {et~al.}(1969)\citenamefont {Michel},
  \citenamefont {Moreau}, \citenamefont {Achenbach}, \citenamefont {Gerson},\
  and\ \citenamefont {James}}]{MMA69}%
  \BibitemOpen
  \bibfield  {author} {\bibinfo {author} {\bibfnamefont {C.}~\bibnamefont
  {Michel}}, \bibinfo {author} {\bibfnamefont {J.~M.}\ \bibnamefont {Moreau}},
  \bibinfo {author} {\bibfnamefont {G.~D.}\ \bibnamefont {Achenbach}}, \bibinfo
  {author} {\bibfnamefont {R.}~\bibnamefont {Gerson}}, \ and\ \bibinfo {author}
  {\bibfnamefont {W.~J.}\ \bibnamefont {James}},\ }\href@noop {} {\bibfield
  {journal} {\bibinfo  {journal} {Solid State Commun.}\ }\textbf {\bibinfo
  {volume} {7}},\ \bibinfo {pages} {865} (\bibinfo {year} {1969})}\BibitemShut
  {NoStop}%
\bibitem [{\citenamefont {Cordero}\ \emph {et~al.}(2009)\citenamefont
  {Cordero}, \citenamefont {Bella}, \citenamefont {Corvasce}, \citenamefont
  {Latino},\ and\ \citenamefont {Morbidini}}]{135}%
  \BibitemOpen
  \bibfield  {author} {\bibinfo {author} {\bibfnamefont {F.}~\bibnamefont
  {Cordero}}, \bibinfo {author} {\bibfnamefont {L.~D.}\ \bibnamefont {Bella}},
  \bibinfo {author} {\bibfnamefont {F.}~\bibnamefont {Corvasce}}, \bibinfo
  {author} {\bibfnamefont {P.~M.}\ \bibnamefont {Latino}}, \ and\ \bibinfo
  {author} {\bibfnamefont {A.}~\bibnamefont {Morbidini}},\ }\href@noop {}
  {\bibfield  {journal} {\bibinfo  {journal} {Meas. Sci. Technol.}\ }\textbf
  {\bibinfo {volume} {20}},\ \bibinfo {pages} {015702} (\bibinfo {year}
  {2009})}\BibitemShut {NoStop}%
\bibitem [{\citenamefont {Nowick}\ and\ \citenamefont {Berry}(1972)}]{NB72}%
  \BibitemOpen
  \bibfield  {author} {\bibinfo {author} {\bibfnamefont {A.~S.}\ \bibnamefont
  {Nowick}}\ and\ \bibinfo {author} {\bibfnamefont {B.~S.}\ \bibnamefont
  {Berry}},\ }\href@noop {} {\emph {\bibinfo {title} {Anelastic Relaxation in
  Crystalline Solids}}}\ (\bibinfo  {publisher} {Academic Press},\ \bibinfo
  {address} {New York},\ \bibinfo {year} {1972})\BibitemShut {NoStop}%
\bibitem [{\citenamefont {Cordero}\ \emph {et~al.}(2016)\citenamefont
  {Cordero}, \citenamefont {Craciun}, \citenamefont {Trequattrini},\ and\
  \citenamefont {Galassi}}]{CCT16}%
  \BibitemOpen
  \bibfield  {author} {\bibinfo {author} {\bibfnamefont {F.}~\bibnamefont
  {Cordero}}, \bibinfo {author} {\bibfnamefont {F.}~\bibnamefont {Craciun}},
  \bibinfo {author} {\bibfnamefont {F.}~\bibnamefont {Trequattrini}}, \ and\
  \bibinfo {author} {\bibfnamefont {C.}~\bibnamefont {Galassi}},\ }\href@noop
  {} {\bibfield  {journal} {\bibinfo  {journal} {arXiv:1602.02799}\ } (\bibinfo
  {year} {2016})}\BibitemShut {NoStop}%
\bibitem [{\citenamefont {Cordero}\ \emph
  {et~al.}(2013{\natexlab{b}})\citenamefont {Cordero}, \citenamefont
  {Trequattrini}, \citenamefont {Craciun},\ and\ \citenamefont
  {Galassi}}]{CTC13}%
  \BibitemOpen
  \bibfield  {author} {\bibinfo {author} {\bibfnamefont {F.}~\bibnamefont
  {Cordero}}, \bibinfo {author} {\bibfnamefont {F.}~\bibnamefont
  {Trequattrini}}, \bibinfo {author} {\bibfnamefont {F.}~\bibnamefont
  {Craciun}}, \ and\ \bibinfo {author} {\bibfnamefont {C.}~\bibnamefont
  {Galassi}},\ }\href@noop {} {\bibfield  {journal} {\bibinfo  {journal} {Phys.
  Rev. B}\ }\textbf {\bibinfo {volume} {87}},\ \bibinfo {pages} {094108}
  (\bibinfo {year} {2013}{\natexlab{b}})}\BibitemShut {NoStop}%
\bibitem [{\citenamefont {Ishchuk}\ and\ \citenamefont {Sobolev}(2015)}]{IS15}%
  \BibitemOpen
  \bibfield  {author} {\bibinfo {author} {\bibfnamefont {V.~M.}\ \bibnamefont
  {Ishchuk}}\ and\ \bibinfo {author} {\bibfnamefont {V.~L.}\ \bibnamefont
  {Sobolev}},\ }\href@noop {} {\bibfield  {journal} {\bibinfo  {journal} {J.
  Surf. Interf. Mater.}\ }\textbf {\bibinfo {volume} {3}},\ \bibinfo {pages}
  {35} (\bibinfo {year} {2015})}\BibitemShut {NoStop}%
\bibitem [{\citenamefont {Dai}\ and\ \citenamefont {Viehland}(1994)}]{DV94}%
  \BibitemOpen
  \bibfield  {author} {\bibinfo {author} {\bibfnamefont {X.}~\bibnamefont
  {Dai}}\ and\ \bibinfo {author} {\bibfnamefont {D.}~\bibnamefont {Viehland}},\
  }\href@noop {} {\bibfield  {journal} {\bibinfo  {journal} {J. Appl. Phys.}\
  }\textbf {\bibinfo {volume} {76}},\ \bibinfo {pages} {3701} (\bibinfo {year}
  {1994})}\BibitemShut {NoStop}%
\bibitem [{\citenamefont {Jiang}\ \emph {et~al.}(1994)\citenamefont {Jiang},
  \citenamefont {Subbarao},\ and\ \citenamefont {Cross}}]{JSC94}%
  \BibitemOpen
  \bibfield  {author} {\bibinfo {author} {\bibfnamefont {Q.~Y.}\ \bibnamefont
  {Jiang}}, \bibinfo {author} {\bibfnamefont {E.~C.}\ \bibnamefont {Subbarao}},
  \ and\ \bibinfo {author} {\bibfnamefont {L.~E.}\ \bibnamefont {Cross}},\
  }\href@noop {} {\bibfield  {journal} {\bibinfo  {journal} {J. Appl. Phys.}\
  }\textbf {\bibinfo {volume} {75}},\ \bibinfo {pages} {7433} (\bibinfo {year}
  {1994})}\BibitemShut {NoStop}%
\bibitem [{\citenamefont {Zhou}\ \emph {et~al.}(2004)\citenamefont {Zhou},
  \citenamefont {Zimmermann}, \citenamefont {Zeng},\ and\ \citenamefont
  {Aldinger}}]{ZZZ04}%
  \BibitemOpen
  \bibfield  {author} {\bibinfo {author} {\bibfnamefont {L.}~\bibnamefont
  {Zhou}}, \bibinfo {author} {\bibfnamefont {A.}~\bibnamefont {Zimmermann}},
  \bibinfo {author} {\bibfnamefont {Y.~P.}\ \bibnamefont {Zeng}}, \ and\
  \bibinfo {author} {\bibfnamefont {F.}~\bibnamefont {Aldinger}},\ }\href@noop
  {} {\bibfield  {journal} {\bibinfo  {journal} {J. Am. Ceram. Soc.}\ }\textbf
  {\bibinfo {volume} {87}},\ \bibinfo {pages} {1591} (\bibinfo {year}
  {2004})}\BibitemShut {NoStop}%
\bibitem [{\citenamefont {Genenko}(2008)}]{Gen08}%
  \BibitemOpen
  \bibfield  {author} {\bibinfo {author} {\bibfnamefont {Y.~A.}\ \bibnamefont
  {Genenko}},\ }\href@noop {} {\bibfield  {journal} {\bibinfo  {journal} {Phys.
  Rev. B}\ }\textbf {\bibinfo {volume} {78}},\ \bibinfo {pages} {214103}
  (\bibinfo {year} {2008})}\BibitemShut {NoStop}%
\bibitem [{\citenamefont {Genenko}\ and\ \citenamefont {Lupascu}(2007)}]{GL07}%
  \BibitemOpen
  \bibfield  {author} {\bibinfo {author} {\bibfnamefont {Y.~A.}\ \bibnamefont
  {Genenko}}\ and\ \bibinfo {author} {\bibfnamefont {D.~C.}\ \bibnamefont
  {Lupascu}},\ }\href@noop {} {\bibfield  {journal} {\bibinfo  {journal} {Phys.
  Rev. B}\ }\textbf {\bibinfo {volume} {75}},\ \bibinfo {pages} {184107}
  (\bibinfo {year} {2007})}\BibitemShut {NoStop}%
\bibitem [{\citenamefont {Woodward}\ \emph {et~al.}(2004)\citenamefont
  {Woodward}, \citenamefont {Reaney}, \citenamefont {Yang}, \citenamefont
  {Dickey},\ and\ \citenamefont {Randall}}]{WRY04}%
  \BibitemOpen
  \bibfield  {author} {\bibinfo {author} {\bibfnamefont {D.~I.}\ \bibnamefont
  {Woodward}}, \bibinfo {author} {\bibfnamefont {I.~M.}\ \bibnamefont
  {Reaney}}, \bibinfo {author} {\bibfnamefont {G.~Y.}\ \bibnamefont {Yang}},
  \bibinfo {author} {\bibfnamefont {E.~C.}\ \bibnamefont {Dickey}}, \ and\
  \bibinfo {author} {\bibfnamefont {C.~A.}\ \bibnamefont {Randall}},\
  }\href@noop {} {\bibfield  {journal} {\bibinfo  {journal} {Appl. Phys.
  Lett.}\ }\textbf {\bibinfo {volume} {84}},\ \bibinfo {pages} {4650} (\bibinfo
  {year} {2004})}\BibitemShut {NoStop}%
\bibitem [{\citenamefont {Damjanovic}(1998)}]{Dam98}%
  \BibitemOpen
  \bibfield  {author} {\bibinfo {author} {\bibfnamefont {D.}~\bibnamefont
  {Damjanovic}},\ }\href@noop {} {\bibfield  {journal} {\bibinfo  {journal}
  {Rep. Prog. Phys.}\ }\textbf {\bibinfo {volume} {61}},\ \bibinfo {pages}
  {1267} (\bibinfo {year} {1998})}\BibitemShut {NoStop}%
\bibitem [{\citenamefont {Cordero}(2007)}]{130}%
  \BibitemOpen
  \bibfield  {author} {\bibinfo {author} {\bibfnamefont {F.}~\bibnamefont
  {Cordero}},\ }\href@noop {} {\bibfield  {journal} {\bibinfo  {journal} {Phys.
  Rev. B}\ }\textbf {\bibinfo {volume} {76}},\ \bibinfo {pages} {172106}
  (\bibinfo {year} {2007})}\BibitemShut {NoStop}%
\bibitem [{\citenamefont {Zhou}\ \emph {et~al.}(2006)\citenamefont {Zhou},
  \citenamefont {Zuo}, \citenamefont {Rixecker}, \citenamefont {Zimmermann},
  \citenamefont {Utschig},\ and\ \citenamefont {Aldinger}}]{ZZR06}%
  \BibitemOpen
  \bibfield  {author} {\bibinfo {author} {\bibfnamefont {L.}~\bibnamefont
  {Zhou}}, \bibinfo {author} {\bibfnamefont {R.~Z.}\ \bibnamefont {Zuo}},
  \bibinfo {author} {\bibfnamefont {G.}~\bibnamefont {Rixecker}}, \bibinfo
  {author} {\bibfnamefont {A.}~\bibnamefont {Zimmermann}}, \bibinfo {author}
  {\bibfnamefont {T.}~\bibnamefont {Utschig}}, \ and\ \bibinfo {author}
  {\bibfnamefont {F.}~\bibnamefont {Aldinger}},\ }\href@noop {} {\bibfield
  {journal} {\bibinfo  {journal} {J. Appl. Phys.}\ }\textbf {\bibinfo {volume}
  {99}},\ \bibinfo {pages} {044102} (\bibinfo {year} {2006})}\BibitemShut
  {NoStop}%
\bibitem [{\citenamefont {Uchino}(1998)}]{Uch98}%
  \BibitemOpen
  \bibfield  {author} {\bibinfo {author} {\bibfnamefont {K.}~\bibnamefont
  {Uchino}},\ }\href@noop {} {\bibfield  {journal} {\bibinfo  {journal} {Acta
  Mater.}\ }\textbf {\bibinfo {volume} {46}},\ \bibinfo {pages} {3745}
  (\bibinfo {year} {1998})}\BibitemShut {NoStop}%
\bibitem [{\citenamefont {Tan}\ \emph {et~al.}(2014)\citenamefont {Tan},
  \citenamefont {Young}, \citenamefont {Seo}, \citenamefont {Zhang},
  \citenamefont {Hong},\ and\ \citenamefont {Webber}}]{TYS14}%
  \BibitemOpen
  \bibfield  {author} {\bibinfo {author} {\bibfnamefont {X.}~\bibnamefont
  {Tan}}, \bibinfo {author} {\bibfnamefont {S.~E.}\ \bibnamefont {Young}},
  \bibinfo {author} {\bibfnamefont {Y.~H.}\ \bibnamefont {Seo}}, \bibinfo
  {author} {\bibfnamefont {J.~Y.}\ \bibnamefont {Zhang}}, \bibinfo {author}
  {\bibfnamefont {W.}~\bibnamefont {Hong}}, \ and\ \bibinfo {author}
  {\bibfnamefont {K.~G.}\ \bibnamefont {Webber}},\ }\href@noop {} {\bibfield
  {journal} {\bibinfo  {journal} {Acta Mater.}\ }\textbf {\bibinfo {volume}
  {62}},\ \bibinfo {pages} {114} (\bibinfo {year} {2014})}\BibitemShut
  {NoStop}%
\end{thebibliography}%

\end{document}